\documentclass[fleqn,usenatbib]{mnras}

\usepackage{newtxtext,newtxmath}

\usepackage[T1]{fontenc}
\usepackage{listings}
\DeclareRobustCommand{\VAN}[3]{#2}
\let\VANthebibliography\thebibliography
\def\thebibliography{\DeclareRobustCommand{\VAN}[3]{##3}\VANthebibliography}

\usepackage{graphicx}	
\usepackage{amsmath}	
\usepackage{mathtools}
\usepackage{gensymb}
\usepackage{CJKutf8}
\usepackage{longtable,booktabs,etoolbox,multirow,tabularx}
\usepackage{algpseudocode}
\usepackage{float}
\usepackage{subcaption}
\usepackage{soul}
\usepackage{hyperref}

\def\msun{M_\odot}

\newcommand*\diff{\mathop{}\!\mathrm{d}}

\title[NLR of GSN 069]{Constraints on the narrow-line region of the X-ray quasi-periodic eruption source GSN\,069}

\author[Patra et al.]{
Kishore C. Patra$^{1, \dagger,}$\thanks{E-mail: kcpatra@berkeley.edu}, 
Wenbin Lu$^{1}$,
Yilun Ma$^{2}$,
Eliot Quataert$^{2}$,  
Giovanni Miniutti$^{3}$, 
Marco Chiaberge$^{4, 5}$, \newauthor
Alexei V. Filippenko$^{1}$ \\
$^{1}$Department of Astronomy, University of California, Berkeley, CA 94720-3411, USA\\
$^{2}$Department of Astrophysical Sciences, Princeton University, Princeton, NJ 08544, USA\\
$^{3}$Centro de Astrobiolog\'ia (CAB), CSIC-INTA, Camino Bajo del Castillo s/n, ESAC campus, 28692 Villanueva de la Ca\~nada, Madrid, Spain \\
$^{4}$ Space Telescope Science Institute for the European Space Agency (ESA), ESA Office, 3700 San Martin Drive, Baltimore, MD 21218, USA \\
$^{5}$ The William H. Miller III Department of Physics and Astronomy, Johns Hopkins University, Baltimore, MD 21218, USA \\
$^{\dagger}$Nagaraj-Noll-Otellini Graduate Fellow \\
}

\date{Accepted XXX. Received YYY; in original form ZZZ}

\pubyear{2023}

\begin{document}
\label{firstpage}
\pagerange{\pageref{firstpage}--\pageref{lastpage}}
\maketitle

\begin{abstract}
The origins of quasi-periodic eruptions (QPEs) are poorly understood, although most theoretical explanations invoke an accretion disk around a supermassive black hole. The gas and stellar environments in the galactic nuclei of these sources are also poorly constrained. In this paper, we present an analysis of archival {\it Hubble Space Telescope (HST)} images to study the narrow-line [O\,III] emission in the QPE source GSN\,069. We find strong evidence for a compact nuclear [O\,III] emission region of size $\lesssim 35$\,pc, overlaid on top of extended [O\,III] emission up to 2\,kpc away from the nucleus. The age of the accretion system is estimated to be between 10 and 100\,yr. 
The [O\,III] luminosity of the compact region was measured to be $(2.1 \pm 0.3) \times 10^{40}\,\rm erg\,s^{-1}$. Based on \textsc{Cloudy} simulations, we constrain that the [O\,III] emitting gas has a hydrogen number density in the range $5 \times 10^{3} < n_{\rm H} \lesssim 10^{8}\,\rm cm^{-3}$ and volume filling factor $f_{\rm V} < 2 \times 10^{-3}$. We suggest that the dense gas in the nuclear region of GSN\,069 originates from molecular clouds (with total mass $\gtrsim 3 \times 10^{3}\,M_{\odot}$), which are freshly ionised by the soft X-ray photons from the accretion disk.
We predict possible evolution of the compact narrow-line region on emission-line diagnostic diagrams, and hence future {\it HST} or integral-field unit observations can be used to further pin down the age of this puzzling system.

\end{abstract}









\section{Introduction}
\label{sec:introduction}

Quasi-periodic eruptions (QPEs) are a new class of X-ray sources identified by wide-field X-ray surveys. These enigmatic objects exhibit distinctive characteristics, featuring bursts of bright X-ray emission
superimposed on top of quiescent emission consistent with steady-state accretion around a supermassive black hole (SMBH). The quasi-periodic recurrence times range from a few hours up to a day in the currently known sample. In the X-ray band, the spectra of QPEs are thermal-like, with higher temperatures ($k_{\rm B}T \approx 100$\,eV) during the eruption phase and lower temperatures ($k_{\rm B}T \approx 50$\,eV) in quiescence. The flare amplitude can reach 100 times that of the quiescent phase at the highest X-ray energies. QPEs are predominantly detected in the nuclei of low-mass galaxies (stellar mass $M_* \approx 10^9\,\msun$), implying that these sources are likely associated with relatively low-mass SMBHs. 
To date, only 6 QPE sources or candidates have been discovered: GSN\,069 \citep{Miniutti_2013, Shu_2018_GSN069, Miniutti_2019}, RX\,J1301.9+2747 \citep{Sun_2013_RXJ, Shu_2017_RXJ, Giustini_2020}, eRO-QPE1 and eRO-QPE2 \citep{Arcodia_2021}, XMMSL1~J024916.6-04124 \citep{Chakraborty_2021}, and ``Tormund'' \citep{Quintin_2023}. Given the small number of known QPEs and sparse temporal/wavelength coverage of them, there is limited knowledge regarding their origin(s), emission processes, lifetime, and periodicity, making them a novel and exciting prospect for understanding the physical processes near these low-mass SMBHs.

Several theoretical frameworks have been proposed to explain the origin of QPEs, spanning concepts such as disk instabilities \citep{Raj_2021, Pan_2022, Pan_2023, Kaur_2023}, mass transfer from various configurations of orbiting bodies \citep{Chen_2022, King_2020, King_2022, Krolik_2022, Metzger_2022, Wang_2022, Zhao_2022, Linial_Sari_2023}, and interactions between a secondary orbiting object and the accretion disk around the primary SMBH \citep{Sukov_2021, Xian_2021, Lu_Quataert_2023, Franchini_2023, Linial_2023}. A connection between QPEs and tidal-disruption events (TDEs) has also been proposed (see, e.g., \citealp{Xian_2021, Wang_2022, King_2020, Linial_2023}). Apparently, two of the currently known QPE sources/candidates might be associated with X-ray-selected TDEs (GSN\,069 and XMMSL1; \citealp{Miniutti_2023A&A, Chakraborty_2021}). A common theme among all the aforementioned models is an accretion disk, whose X-ray emission has been detected (during the quiescence phase) in at least some of the QPE sources. There is some indication that these accretion disks may be compact in size, as inferred from the lack of broad optical emission lines and near-infrared (IR) dust emission typically associated with active galactic nuclei (AGNs). However, the disk properties (e.g., accretion rate, mass feeding source, and radial structure) are currently poorly constrained. Much also remains unknown about the gas and stellar environments in these galactic nuclei. 

In this paper, we present the analysis of {\it Hubble Space Telescope (HST)} images to study the narrow-line region in the nucleus of the first-ever QPE source GSN\,069 [$\alpha (\rm J2000) = 01^{\rm hr} 19^{\rm m} 08.61^{\rm s}; ~ \delta (\rm J2000) = -34^{\degree} 11^{\prime} 30.13^{\prime \prime}$; $z = 0.018$]. GSN\,069 was first discovered in July 2010 during an {\it XMM-Newton} slew \citep{Saxton_2011}, exhibiting a flux over 240 times higher than previous {\it Roentgen Satellite (ROSAT)} observations from 1994 \citep{Miniutti_2013}. Subsequent observations with the {\it Swift} X-Ray Telescope (XRT) indicated a relatively steady X-ray flux for around a year \citep{Miniutti_2013}. {\it Swift} and {\it XMM-Newton} further showed a gradual decline in flux over the next decade \citep{Shu_2018_GSN069}. Notably, QPEs --- X-ray flares with high amplitude and short duration, repeating approximately every 9\,hr --- were observed during {\it XMM-Newton} and {\it Chandra} observations between December 2018 and February 2019 \citep{Miniutti_2019}. Mysteriously, the QPE phenomenon in this source goes on and off together with secular evolution of the quiescent X-ray flux \citep{Miniutti_2023, Miniutti_2023A&A}.

This paper is organized as follows. Section \ref{sec:images} provides details on the archival {\it HST} images and the process of creating a continuum-subtracted image for the [O\,III] line emission. In Section \ref{sec:photometry}, we describe our photometry model along with the results. Subsequently, we present our \textsc{Cloudy} modeling along with constraints on the [O\,III]-emitting gas in Section \ref{sec:cloudy}. We discuss our findings and conclusions in Sections \ref{sec:discussion} and \ref{sec:conclusion}, respectively.

\section{HST Images}
\label{sec:images}

\begin{figure} 
    \centering
    \includegraphics[width=.5\textwidth]{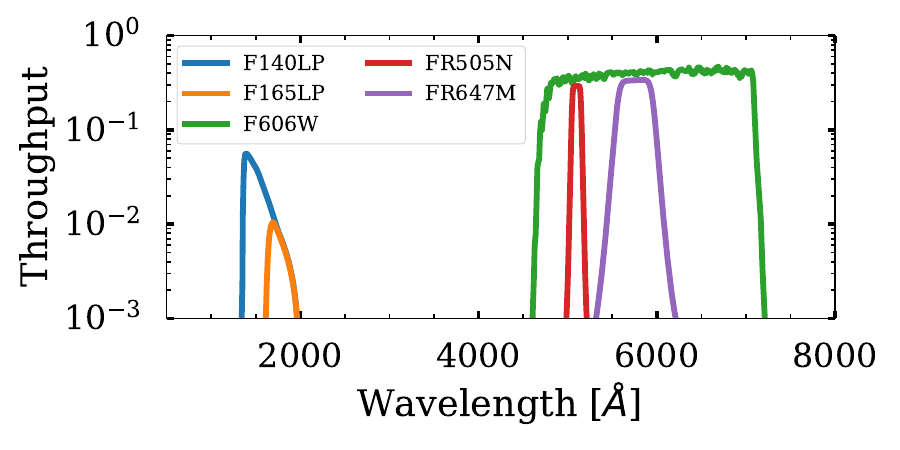}
    \caption{The throughput functions of the {\it HST} filters in which the images of GSN\,069 were obtained.}
    \label{fig:hst_filters}
\end{figure}

Archival {\it HST} images of GSN\,069 were retrieved from the Mikulski Archive for Space Telescopes (MAST). These images were originally obtained in {\it HST} Cycle 27 (Program ID 16062; PI: G. Miniutti) in August and November 2020 with the goal, among others, to study the spatially-resolved structure of the narrow-line region. Five images were obtained with the Advanced Camera for Surveys (ACS) over two {\it HST} orbits: two far-ultraviolet (UV) band (F140LP, F165LP) images, one narrow-band optical image to capture the [O\,III] emission (FR505N), a medium-band image (FR647M) for the continuum near the [O\,III] line, and a wide-band optical image (F606W) to study the spatially-resolved stellar structure. Table \ref{tab:observations} provides a summary of the {\it HST} observations including the filters, observation dates, exposure times, and the central wavelengths $\lambda_{\rm centre}$ and  full width at half maximum intensity (FWHM) of the filter throughput functions. The filter functions are shown in Figure \ref{fig:hst_filters}. The middle segment of the ramp filter FR505N with central wavelength $5099$\,\AA\ was used to capture the [O\,III] $\lambda 4959$ and $\lambda 5007$ lines which were redshifted to $5046$\,\AA\ and $5097$\,\AA\ (respectively) by the recession of the host galaxy GSN\,069 at redshift $z = 0.018$. Standard {\it HST} reductions and calibrations were carried out with \textit{AstroDrizzle}. 

\begin{table*}
    \centering
    \begin{tabular}{l  c  c  c  c  c}
     \hline
    Filter	& UT Date (yyyy-mm-dd) & $\lambda_{\rm centre}$ [\AA] & FWHM [\AA]& $T_{\rm exp}$ [s] & Scale [kpc/$''$]\\
      \hline\hline
      F140LP & 2020-11-19 & 1476 & 226 & 1000 & 0.19 \\
      F165LP & 2020-11-19 & 1740 & 200 & 1000 & 0.19 \\
      FR505N & 2020-08-16 & 5099 & 105 & 700 & 0.38\\
      FR647M & 2020-08-16 & 5757 & 499 & 160 & 0.38\\
      F606W  & 2020-08-16 & 5962 & 2253 & 800 & 0.38\\
    \hline
    \end{tabular}
    \caption{Summary of {\it HST} observations of GSN 069.}
    \label{tab:observations}
\end{table*}

The FR505N image contains both the [O\,III] line and continuum fluxes, and we first created the continuum-subtracted [O\,III] image as follows. Since the available images lacked a sufficient number of resolved stars, registering the FR505N (on-band) and FR647M (continuum) images based on stars was not feasible. Instead, we computed the photometric centroids of the galactic nuclei in both the on-band and continuum images. Subsequently, we generated a $100\times100$ pixel ($\sim 2$\,kpc $\times\, 2$\,kpc in physical size) cutout centred around each centroid. Despite the centroid's accuracy being better than an {\it HST} pixel, subtraction between the two images is no better than one pixel. To address this challenge, we employed a strategy of supersampling --- we increased the resolution of both cutout images by a factor of 5 through polynomial interpolation between adjacent pixels. For each image, the pixel counts in units of $\rm electron\,s^{-1}$ were converted to flux density in units of $\rm erg\,s^{-1}\,cm^{-2}$\,\AA$^{-1}$ using the PHOTFLAM keyword in the image headers. Subsequently, we subtracted the supersampled continuum cutout from the supersampled on-band cutout, both expressed in flux-density units.

\begin{figure*}
\centering
  \includegraphics[width=\linewidth]{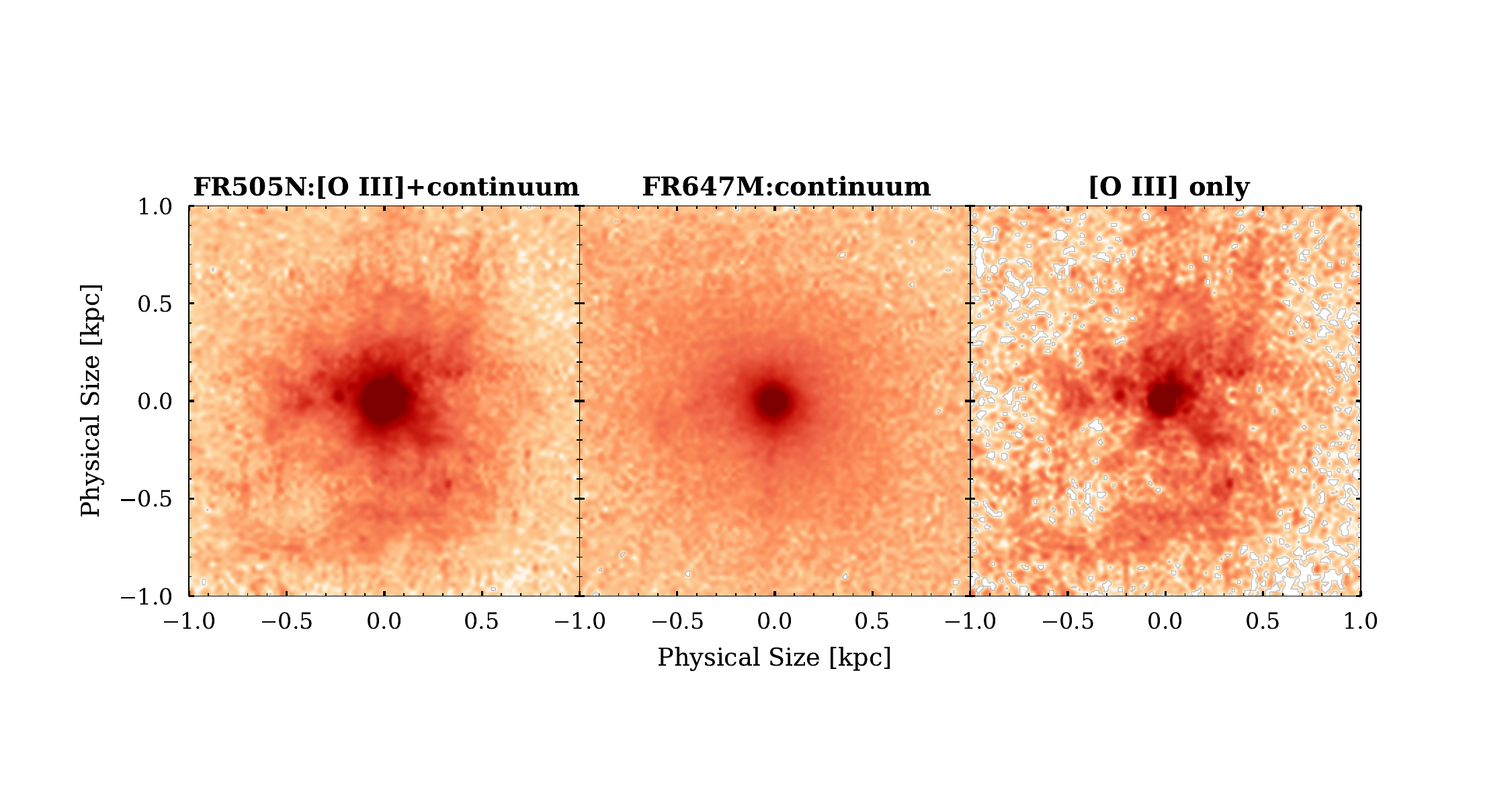}
  \caption{The left panel shows the {\it HST} image in the narrow-band filter FR505N which captures [O\,III] emission at 4958.9\,\AA\ and 5006.9\,\AA, along with the continuum emission. The middle panel shows the continuum image in filter FR647M. The right panel is the continuum-subtracted image of GSN\,069.} 
  \label{fig:contfree}
\end{figure*}

Figure \ref{fig:contfree} presents the resultant continuum-free image displaying the [O\,III] emission. It is apparent that the surface brightness profile of [O\,III] emission has a cusp structure near the nucleus and a number of elongated structures away from the nucleus. Based on the [O\,III] image alone, it is not clear whether these structures reflect the spatial distribution of dense gas or sources of ionising photons or both. To gauge the systematic uncertainty associated with the [O\,III] flux due to the process of image subtraction, a deliberate misalignment of the images in random directions, each by a slight amount (1-2 supersampled pixels), was undertaken. This procedure allowed us to determine the average systematic difference in [O\,III] flux as a function of the distance from the centre. The resultant systematic-error function was then combined with the statistical error to estimate the comprehensive uncertainty linked to the flux measurement.

\begin{figure*}
  \centering
  \includegraphics[width=\linewidth]{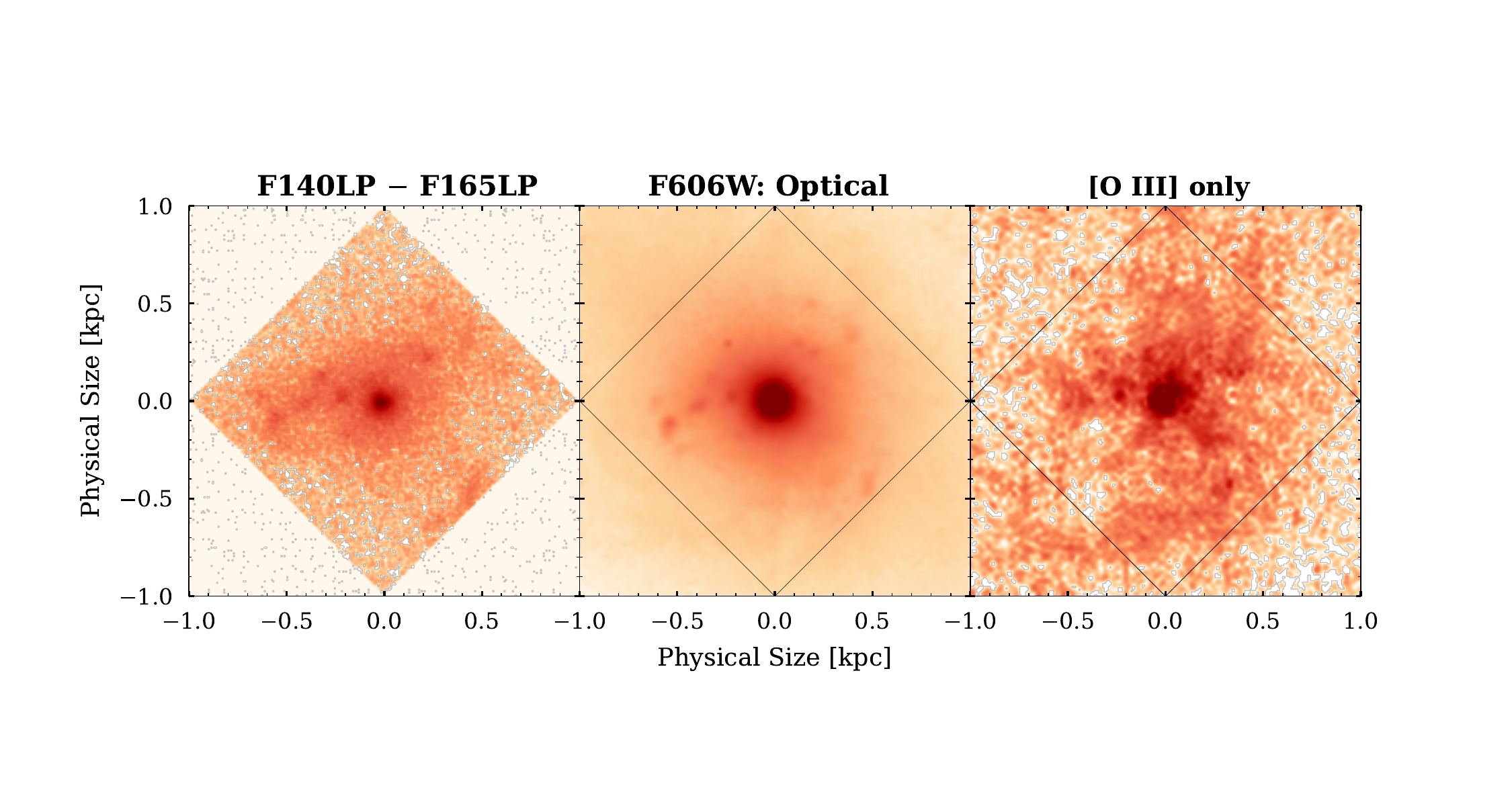}
  \caption{The left panel shows the far-UV image create by subtracting F165LP from F140LP. The middle panel displays the broad-band F606W optical image. The continuum-free [O\,III] emission image is reproduced for comparison in the right panel.}
  \label{fig:FUVoptical}
\end{figure*}

In Figure \ref{fig:FUVoptical}, we show the far-UV (F140LP$-$F165LP)\footnote{The {\it HST} Solar Blind Camera (SBC) has a known problem of red leak for wavelengths $> 2000$\,\AA. To remove the contaminating flux from longer wavelengths, and to get the true UV flux between 1400 and 1600\,\AA, we subtracted the two far-UV images F140LP and F165LP. This works because the throughput function of F165LP is contained entirely within the throughput function of F140LP as can be seen in Figure \ref{fig:hst_filters}.} and broad-band optical (F606W) images for comparison.
From the fact that the elongated structures on larger scales $\gtrsim 0.3\,$kpc in the [O\,III] image do not appear to correspond to any features in the far-UV image (young stars) and the broad-band optical image (young and old stars), we infer that the [O\,III] emission on larger scales is not dominated by the photoionised (H\,II) regions near young stars. 
We leave a detailed analysis of the spectral energy distributions of the nucleus and the surrounding stars to a future work, whereas in this paper we focus on the [O\,III] emission.

\section{Photometry}
\label{sec:photometry}

Aperture photometry was performed using \textsc{Astropy}'s \textsc{Photutils} module \citep{Astropy_collab_2022}. Annular apertures of varying radii were centred on the calculated centroid of the continuum-free [O\,III] image. The median surface flux density $\mu_{\rm \lambda, o} (r)$ was calculated in each annular aperture. 
We modeled the observed [O\,III] image as a convolution of a source image with a point-spread function (PSF),
\begin{equation}
\mu_{\rm \lambda, o} (x,y) = \int \diff x^{\prime} \diff y^{\prime} \mu_{\rm \lambda, s} (x^{\prime},y^{\prime}) K(\Delta r), 
\end{equation}
where $\Delta r = \sqrt{(x-x^{\prime})^{2} + (y-y^{\prime})^{2}}$, $\mu_{\rm \lambda, o} (x,y)$ is the observed flux density at pixel position $(x,y)$, $\mu_{\rm \lambda, s} (x^{\prime},y^{\prime})$ is the flux density of the source, and $K(\Delta r)$ is the PSF, also known as the kernel function in the context of convolution. We used the PSF of the filter F555W which encapsulates the relevant wavelengths of both FR505N and FR647M. Annular photometry was subsequently carried out on the convolved image to get a model surface flux density as a function of radius $r = \sqrt{x^2 + y^2}$. 

\begin{figure*} 
    \centering
    \includegraphics[width=0.8\textwidth,keepaspectratio]{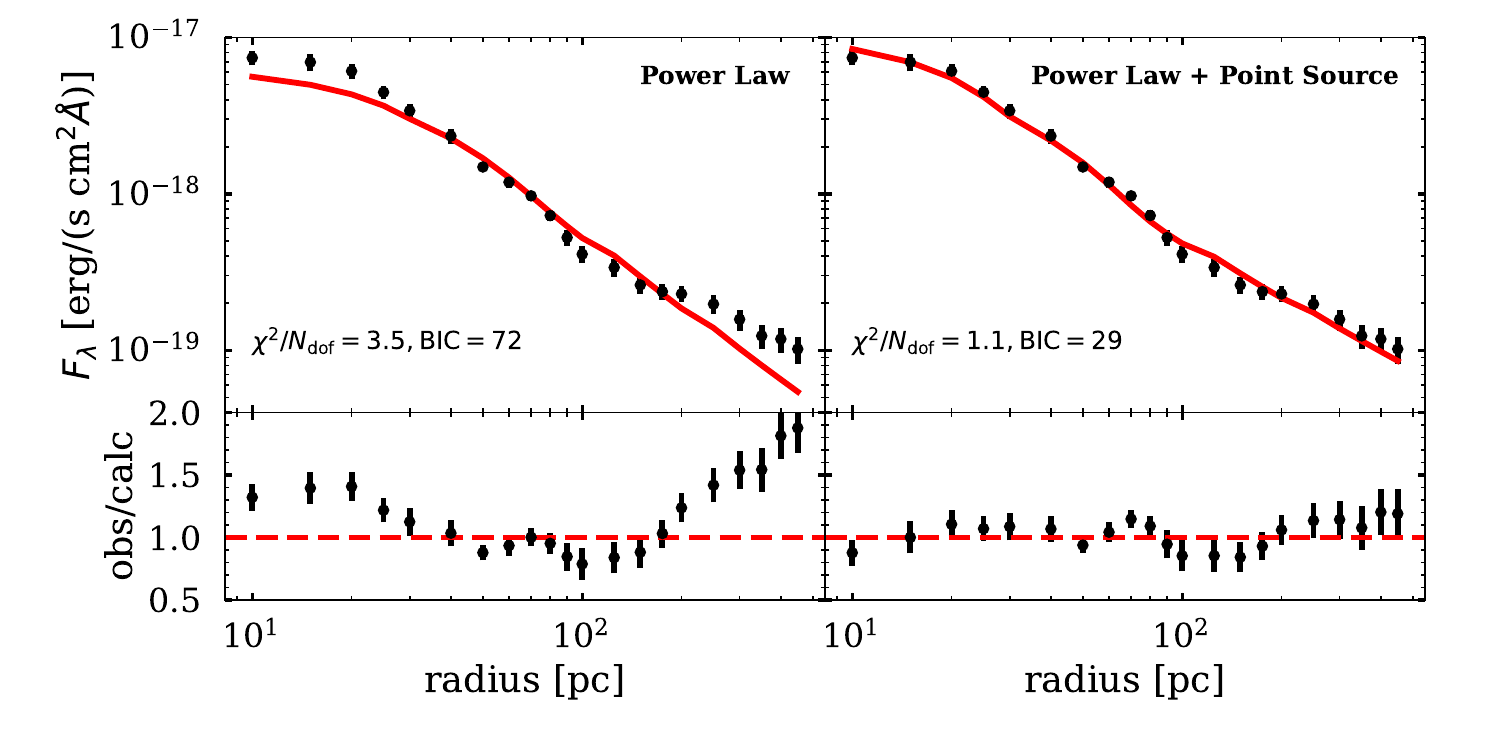}
    \caption{Fitting of photometry models to the observed data. Left panel shows the fitting of a power-law-only model, whereas the right panel shows the fitting of a power-law + point-source model. The bottom panels display observed data-to-best-fit model ratios as residuals.}
    \label{fig:modelfits}
\end{figure*}

\begin{figure*}
\centering
\begin{subfigure}{.4\textwidth}
  \centering
  \includegraphics[width=\linewidth]{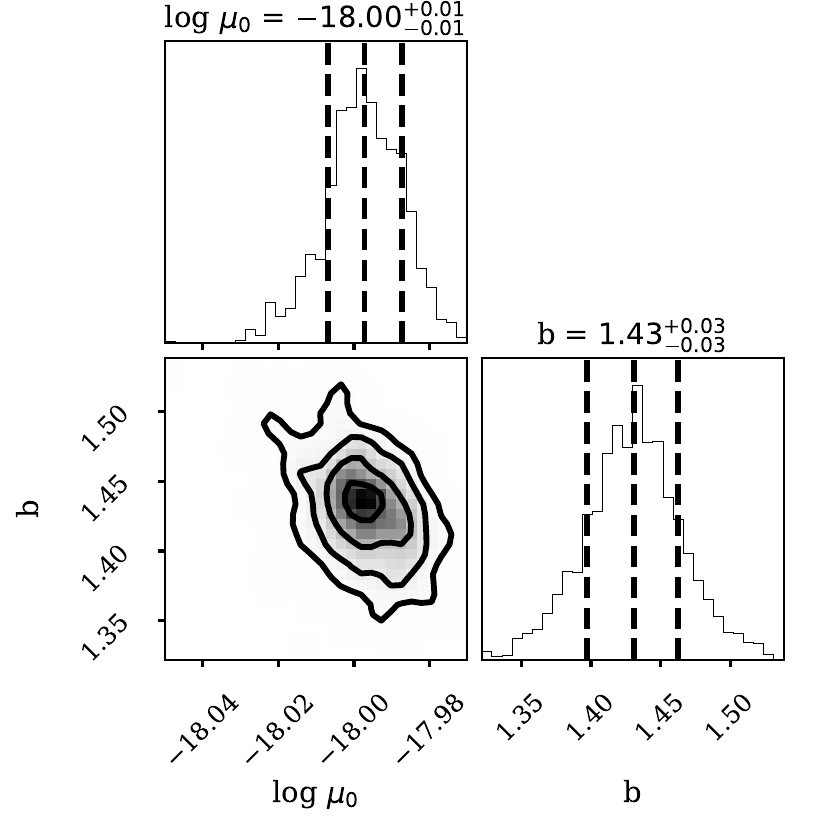}
  \caption{Power law model}
  \label{fig:sub1}
\end{subfigure}%
\begin{subfigure}{.4\textwidth}
  \centering
  \includegraphics[width=\linewidth]{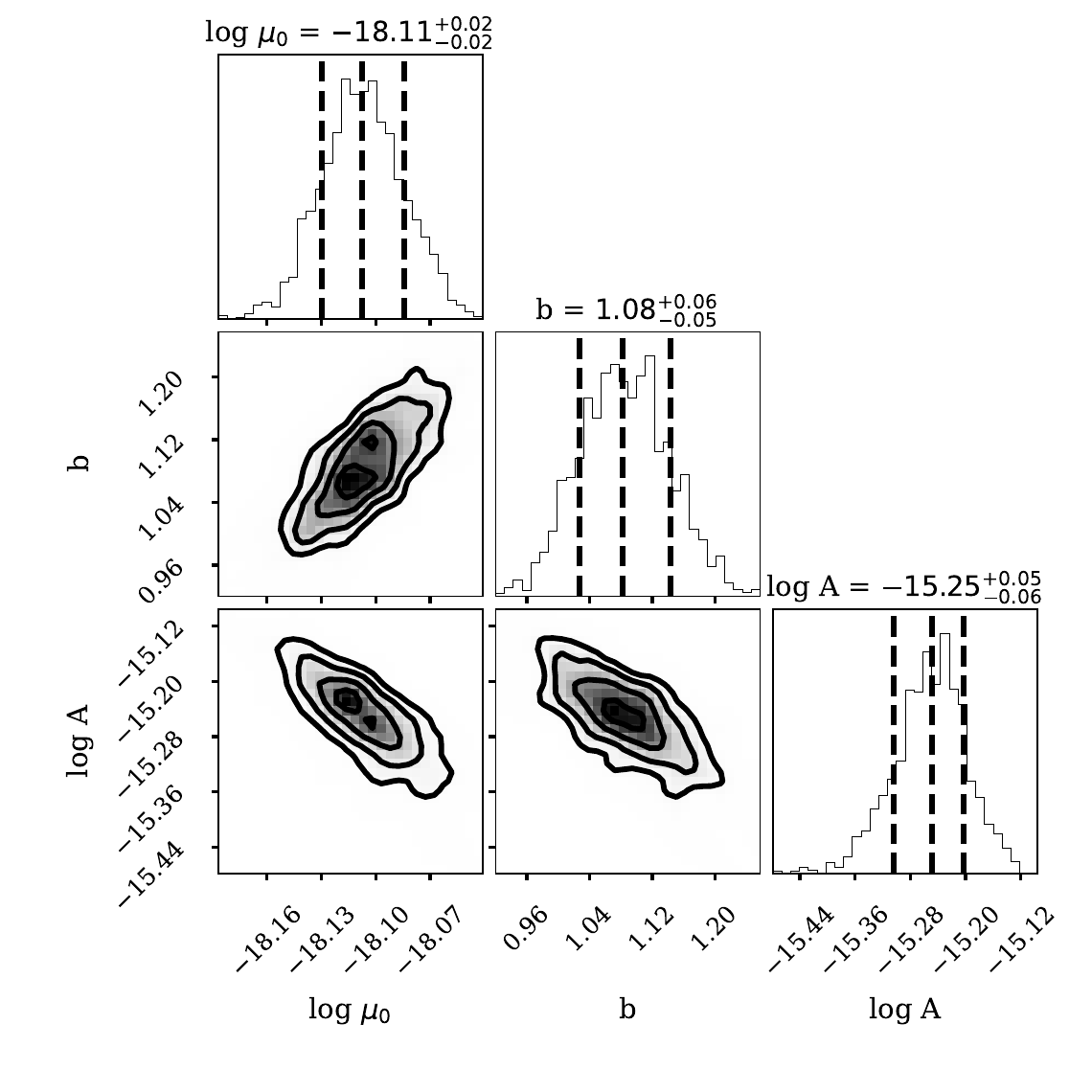}
  \caption{Power-law + point-source model}
  \label{fig:sub2}
\end{subfigure}
\caption{Posterior distributions of the parameters that describe the two models fit to the photometry of the continuum-subtracted [O\,III] image.}
\label{fig:posterior_dist}
\end{figure*}

Two source models were tested, with the first being a power law model given by 
\begin{equation}\label{eqn:powerlaw}
\mu_{\rm s} (r) = \mu_{0} \left( \frac{r}{r_{0}} \right)^{-b}, 
\end{equation}
where $\mu_{0}$ is the surface flux density at a fixed reference radius $r_{0} = 50$\,pc (the results do not depend on this choice), and $b$ is the power-law index. To overcome the numerical singularity at $r = 0$, the value of the central pixel was extrapolated using the gradient at immediately adjacent pixels. The second model was a power-law model plus a point source given by
\begin{equation}\label{eqn:powerlaw+delta}
\mu_{\rm s} (r) = \mu_{0} \left( \frac{r}{r_{0}} \right)^{-b} + A \delta[r], 
\end{equation}
where $A$ is the amplitude (in units of surface flux density) of the discrete unit sample function $\delta[r]$, which is defined as
\[
    \delta[r]= 
\begin{cases}
    1, & \text{if } r = 0\\
    0,              & \text{otherwise.}
\end{cases}
\]
Figure \ref{fig:modelfits} shows the annular photometry along with the best-fit models, whereas Figure \ref{fig:posterior_dist} shows the posterior distributions of the fitted parameters for the two models. Table \ref{tab:photometry_all} displays the measured flux density within circular apertures of various sizes for all {\it HST} images of GSN\,069. We acknowledge that the PSF of F555W is merely an approximation of the PSF of the continuum-free image. However, the fitted parameters in Equations \ref{eqn:powerlaw} and \ref{eqn:powerlaw+delta} exhibited relative robustness when different optical band filters' PSFs were utilised.


The power-law model does not provide a good fit to the observed data, as indicated by a reduced chi-squared of $\chi_{\rm red}^2=3.5$ with 19 degrees of freedom (dof).  
The Bayesian Information Criterion (BIC), defined as BIC $\equiv \chi^2 + k\ln{n}$, where $k$ represents the number of free parameters in the model and $n$ is the quantity of data points, yields a BIC value of 72 for this model. In contrast, the power-law + point-source model demonstrates an excellent fit to the data, with $\chi_{\rm red}^2=1.1$ with 18 degrees of freedom and a BIC of 29. The evidence strongly favours the power law + point-source model over the power-law model. Although we do not exhaust all possible parametrised models with more parameters (e.g., a 4-parameter Gaussian + power-law model, or a 4-parameter broken-power-law model), our preferred power-law + point-source model provides an excellent fit to the data with only three parameters.

We interpret these results as 
being consistent with the existence of a compact, unresolved source of [O\,III] emission within the nucleus of GSN\,069, superimposed onto an extended power-law distribution of [O\,III] emission. The best-fit values for the free parameters in the power-law + point-source model are $\log \mu_{0} = -18.11 \pm 0.02\,\rm erg\,s^{-1}\,cm^{-2}$\,\AA$^{-1}$, $b = 1.08 \pm 0.06$, and $\log A = -15.25 \pm 0.06\,\rm erg\,s^{-1}\,cm^{-2}$\,\AA$^{-1}$. The measured $\log A$ parameter corresponds to a luminosity of the compact [O\,III] source of $(2.1 \pm 0.3) \times 10^{40}\,\rm erg\, s^{-1}$. We discuss the implications of these results in the next section. On the other hand, the best-fit values for the free parameters in the power-law-only model are $\log \mu_{0} = -18.00 \pm 0.01\,\rm erg\,s^{-1}\,cm^{-2}$\,\AA$^{-1}$ and $b = 1.43 \pm 0.03$. 


\begin{table*}
\centering
\begin{tabular}{cc|ccccc|ccccc}
Filter & $\lambda$ & \multicolumn{5}{c|}{$\lambda F_{\lambda}$ ($10^{-12}$\,$\rm erg\,s^{-1}\,cm^{-2}$)} & \multicolumn{5}{c}{Uncertainty ($10^{-15}$\,$\rm erg\,s^{-1}\,cm^{-2}$)} \\
 & (\AA) & $0.1''$ & $0.2''$ &$0.3''$ & $0.4''$ & $1.2''$ & $0.1''$ & $0.2''$ &$0.3''$ & $0.4''$ & $1.2''$ \\
\hline \hline
F140LP & 1475 & 1.196 & 1.444 & 1.465 & 1.433 & 2.175 & 8.641 & 8.548 & 8.157 & 7.768 & 8.320 \\
F165LP & 1740 & 1.333 & 1.537 & 1.545 & 1.508 & 2.232 & 2.255 & 2.211 & 2.122 & 2.027 & 2.164 \\
FR505N & 5099 & 1.623 & 2.103 & 2.413 & 2.646 & 5.389 & 6.841 & 6.947 & 7.263 & 7.530 & 10.206 \\
FR647M & 5757 & 1.559 & 1.921 & 2.177 & 2.362 & 4.212 & 6.767 & 6.620 & 6.865 & 7.068 & 8.974 \\
F606W & 5962 & 1.385 & 1.858 & 2.109 & 2.296 & 4.137 & 1.110 & 1.142 & 1.184 & 1.223 & 1.560 \\
\hline
\end{tabular}
\caption{Flux measurement of all {\it HST} images. The radii in arcseconds of the circular apertures are shown in the header. The uncertainty presented here is the statistical Poisson noise.}
\label{tab:photometry_all}
\end{table*}

\section{\textsc{Cloudy} Simulations}
\label{sec:cloudy}

We used \textsc{Cloudy} \citep{Ferland_2017} --- a spectral synthesis code designed to simulate the conditions of gas-radiation interactions --- to calculate the [O\,III] line emissivity  on a two-dimensional (2D) grid of hydrogen number density and distance from the SMBH, with the goal of constraining the physical properties the [O\,III] emitting gas. We varied the hydrogen number density $n_{\rm H}$ from $3 \times 10^3\,\rm cm^{-3}$ to $10^6\,\rm cm^{-3}$, while varying the distance $r$ from 1\,pc to 100\,pc.
Motivated by X-ray observations in the past two decades \citep[see][for a summary]{Miniutti_2023A&A}, a spectral energy distribution (SED) table was created for a standard multicolour-blackbody accretion-disk model\footnote{The standard multicolour-blackbody disk SED has the asymptotic behaviour of $\nu L_\nu\propto \nu^{4/3}$ at low frequencies $h\nu \ll k_{\rm B}T_{\rm in}$ \citep{Shakura_Sunyaev_1973}. This is only possible if the disk extends to infinity with a constant mass accretion rate at all radii. \citet{Lu_Quataert_2023} pointed out that fresh gas may join the disk at a finite radius $r_*$, which is on the order  of the pericentre radius of the stellar orbit if the disk is fed by a star (or the tidal disruption of a star). In this case, the outer regions of the disk at $r>r_*$ do not have a constant mass-accretion rate, and the disk SED steepens to $\nu L_\nu\propto \nu^{12/7}$. The disk also has an outer edge at radius $r_{\rm out}$ due to the finite lifetime in which the disk can viscously spread, and hence the emission at the lowest frequencies has a Rayleigh-Jeans spectrum $\nu L_\nu\propto \nu^3$. For our purpose here, since most ionising photons always come from near the inner edge of the disk, the systematic errors introduced by the uncertainty in the disk SED at low frequencies are minor, so we assume the standard disk SED in our model. } \citep{Shakura_Sunyaev_1973} with two conditions: (i) the X-ray luminosity between 0.2\,keV and 2\,keV is taken to be $L_{\rm 0.2\mbox{-}2\,keV} = 4 \times 10^{42}\,\rm erg\,s^{-1}$, and (ii) the temperature at the inner edge of the disk $k_{\rm B}T_{\rm in} = 50\,\rm eV$ ($k_{\rm B}$ being the Boltzmann constant). 
The disk SED is assumed to be time-independent, although in reality, the X-ray flux varied by a factor of 2 above and below the time-averaged value taken in our model. The systematic error introduced by this assumption is minor compared to other sources of uncertainties, as we only aim to provide rough constraints on the [O\,III]-emitting gas properties using a simple one-zone model. The above SED table was fed into the \textsc{Cloudy} simulations.

\begin{figure*} 
    \centering
    \includegraphics[width=\textwidth,keepaspectratio]{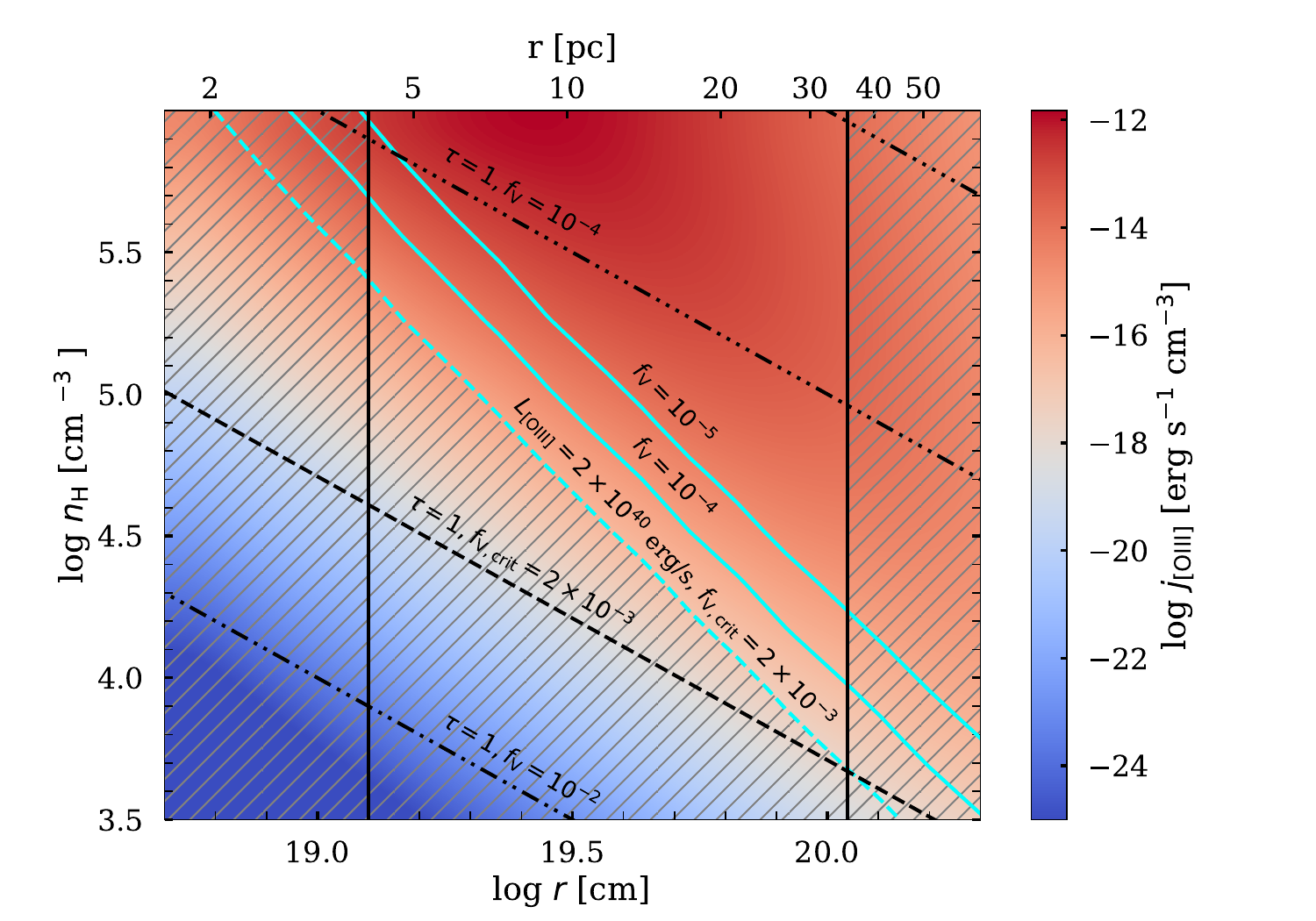}
    \caption{Constraints on the hydrogen number density ($n_{\rm H}$) and the distance from SMBH of the [O\,III]-emitting gas in the nucleus of GSN\,069. The colour bar represents the [O\,III] emissivity calculated by \textsc{Cloudy}. The cyan-coloured bands represent the contours that satisfy the observed [O\,III] luminosity of the unresolved point source for various values of the volume filling factors $f_{\rm V}$. The solid vertical black lines show constraints on the size of the emitting region. The black dash-dotted lines show iso-extinction contours for various values of $f_{\rm V}$. The dash-dotted lines follow this scheme: 2 dots between dashes for $f_{\rm V} = 10^{-2}$, 4 dots between dashes for $f_{\rm V} = 10^{-4}$, and so on. The iso-extinction contour and the [O\,III] luminosity solution corresponding to $f_{\rm V, crit}$ are shown with dashed lines. The disallowed region is hatched with grey lines. 
    }
    \label{fig:emissivity_Oiii}
\end{figure*}

Figure \ref{fig:emissivity_Oiii} displays the volume emissivity of [O\,III], $j_{\rm [O\,III]}$\,($\rm erg\,s^{-1}\,cm^{-3}$), on a 2D grid of hydrogen number density ($n_{\rm H}$) and distance $r$ from the SMBH. The [O\,III] line luminosity is given by
\begin{equation}
L_{\rm [O\,III]}(n_{\rm H}, r, f_{\rm V}) =  (4/3) \pi r^{3} f_{\rm V} j_{\rm [O\,III]},
\label{lum_emissivity}
\end{equation}
where $f_{\rm V}$ is the volume filling factor of the [O\,III]-emitting gas. Here, in our simple one-zone model, $n_{\rm H}$ and $r$ should be understood as the density and distance to the SMBH for the gas that dominates the observed [O\,III] emission. In reality, the [O\,III] luminosity is contributed by different emitting regions with different gas densities, and these variations can be captured by an effective volume filling factor. The cyan lines represent the combinations of $n_{\rm H}$ and $r$ that yield the observed point-source [O\,III] luminosity of $2.1 \times 10^{40}\,\rm erg\,s^{-1}$, for three distinct values of $f_{\rm V}$. The black vertical lines restrict the physical size of the unresolved nuclear [O\,III] source. The disk X-ray emission has been active for at least 13\,yr \citep{Saxton_2011}. During this time, the ionising photons from the accretion disk would have traveled $r_{\rm min} = 4\rm \,pc$, thus setting the lower limit to the size of the nuclear [O\,III] emitting region. On the other hand, the unresolved point-source region can be no larger than the FWHM of the {\it HST} PSF. Consequently, the upper limit for the nuclear [O\,III] region's size is $r_{\rm max}=35\rm\, pc$. 

Additional constraints can be derived from the requirement that the gas in the nucleus of GSN\,069 is optically thin to optical and soft X-ray emission, otherwise the [O\,III] line or the QPE flares will not be detected. Taking a conservative limit for the $V$-band dust extinction $A_{\rm V} < 1 $\,mag, we get a total hydrogen column density $N_{\rm H} < 2 \times 10^{21}\,\rm cm^{-2}$ \citep[see chapter 21.2 of][]{Draine_2011_ISM_book}. This allows us to set iso-extinction contours in the $n_{\rm H}$--$r$ space as 
\begin{equation}\label{extinction_contour}
n_{\rm H} ~r < 2 \times 10^{21}/f_{\rm V}\,\rm cm^{-2}.
\end{equation}
These contours are shown by the dash-dotted lines. A critical volume filling factor $f_{\rm V, crit} = 2 \times 10^{-3}$ was determined such that the corresponding extinction contour and the $L_{\rm [O\,III]}$ solution intersect the $r_{\rm max}$ vertical black line. Any solution above the critical iso-extinction contour for $f_{\rm V, crit}$ is not viable because the [O\,III]-emitting gas would be optically thick, which is inconsistent with observations. Thus, $f_{\rm V} > 2 \times 10^{-3}$ is ruled out. 
The unshaded regions in the $n_{\rm H}$--$r$ space represent the viable combinations of number density and distance to the SMBH for the [O\,III]-emitting gas.

By inspecting Figure \ref{fig:emissivity_Oiii}, we can derive an approximate analytic form of the $L_{\rm [O\,III]}$ solution contours given as 
\begin{equation}\label{analytic_n}
    n \approx 5 \times 10^{5}\,\text{cm}^{-3} ~ \left(\frac{r}{10^{19.1}\,\text{cm}} \right)^{-1.8} ~ \left(\frac{f_{\text{V}}}{10^{-4}} \right)^{-0.3}. 
\end{equation}
The total mass of the [O\,III]-emitting gas is 
\begin{equation}
M_{\text{gas}} = (4/3) \pi r^{3} f_{\rm V} n_{\rm H}  m_{\rm H},
\end{equation} 
where $m_{\rm H}$ is the mass of a hydrogen atom.
It follows that the minimum mass of the [O\,III]-emitting region must correspond to the lower-left corner of the allowed region in $n_{\rm H}$--$r$ space --- that is, where the $L_{\rm [O\,III]}$ contour for $f_{\rm V,crit}$ intersects the $r_{\rm min}$ bound. We found that the total gas mass within the compact nuclear region is at least $3 \times 10^{3}\,M_{\odot}$, a conservative limit since we have excluded diffuse gas that does not contribute significantly to the [O\,III] luminosity.

We also explored the feasibility of establishing an upper limit for $n_{\rm H}$ by comparing the observed H$\alpha$ luminosity with the luminosity produced by the \textsc{Cloudy} simulations. \citet{Wevers_2022} provide the equivalent width (EW) of the $\rm H\alpha$ line but not the actual measurement of the $\rm H\alpha$ flux in a spectrum of GSN\,069 acquired with a $0.7''$-wide slit. To estimate the $\rm H\alpha$ flux, we performed photometry on the HST continuum image (FR647M) with a $0.7''$ aperture and multiplied it with the EW determined by \citet{Wevers_2022}. Consequently, we were able to derive $\rm H\alpha$ luminosity as $L_{\rm H\alpha} \approx 6 \times 10^{40}\,\rm erg\,s^{-1}$. Note that $0.7''$ corresponds to $\sim 300$\,pc in physical size; thus, the $L_{\rm H\alpha}$ value calculated above should be treated as an upper limit on allowed $L_{\rm H\alpha}$ in \textsc{Cloudy} simulations. Unfortunately, the H$\alpha$ luminosity imposes comparable constraints on the $n_{\rm H}$--$r$ space as $L_{\rm [O\,III]}$, yielding no new insights. Consequently, we refrain from discussing $\rm H\alpha$ emission further. Nevertheless, it is important to note that $n_{\rm H}$ cannot be arbitrarily high. Based on \textsc{Cloudy} calculations, we see that for $n_{\rm H} \gtrsim 10^{8}\,\rm cm^{-3}$ , the emissivity of [O\,III] per unit mass rapidly declines to a constant value owing to collisional deexcitation.

Therefore, the hydrogen number density of the gas in the nucleus is constrained to $5 \times 10^{3} < n_{\rm H} \lesssim 10^{8}\,\rm~cm^{-3}$, with a volume filling factor $f_{\rm V} < 2 \times 10^{-3}$. These values of $f_{\rm V}$ and $n_{\rm H}$ are characteristic of dense molecular gas, which are gravitationally bound clouds that are often the sites of star formation before the SMBH becomes active \citep[see Table 1.3 of][]{Draine_2011_ISM_book}. The onset of accretion onto the SMBH then rapidly ionised the surrounding dense gas. Subsequently, the dense gas will have to contend with two opposing pressure forces: the thermal pressure is $P_{\text{gas}} = n k T_{\rm gas}$, whereas the pressure from the accretion disk's radiation field is $P_{\rm rad} = L / 4\pi r^{2} c$. For the region near $r_{\rm min}$, $P_{\text{gas}} \approx 4 \times 10^{-7}\, \rm g\,cm^{-1}\,s^{-2}$ (for a typical value of $T_{\rm gas}\approx 10^4\rm\, K$) and $P_{\text{rad}} \approx 2 \times 10^{-7}\,\rm g\,cm^{-1}\,s^{-2} $. Since $P_{\text{gas}} \approx P_{\text{rad}}$, the dense gas may achieve pressure equilibrium, preventing it from dispersing away (we note that the system is unlikely to be in hydrostatic equilibrium and that the 3D radiation-gas dynamics are much more complicated). Furthermore, even if the gas cloud could expand, it would do so at the speed of sound, $c_{\rm s} = \sqrt{k T_{\rm gas}/m_{\rm p}} \approx 10\,\rm km\,s^{-1}$. For a $100$\,yr-old system (as an example), the gas will only expand by up to $10^{15}$\,cm, which is much smaller than $r_{\rm min}$. We conclude that the [O\,III]-emitting gas may originate from dense molecular clouds in the vicinity of the SMBH at distances $\mathcal{O}(10)\rm\,pc$.


Note that our \textsc{Cloudy} simulations are conducted under the assumption of steady-state ionisation conditions. The time it takes for recombination of O\,III $\rightarrow$ O\,II and H\,II $\rightarrow$ H\,I is roughly given by 
\begin{equation}
\begin{split}
t_{\rm rec, O\,III} \approx 1\,\text{yr} \frac{10^{4}\,\rm cm^{-3}}{n_{\rm H}}, \ \ 
\noindent t_{\rm rec, H\,II} \approx 10\,\text{yr} \frac{10^{4} \,\rm cm^{-3}}{n_{\rm H}}, 
\end{split}
\end{equation}
respectively \citep{Draine_2011_ISM_book}. For permissible values of $n_{\rm H} > 5 \times 10^{3}\,\rm cm^{-3}$, the recombination timescale for O\,III is much shorter than the estimated duration for which the SMBH has been active (10--100\,yr). For most allowed values of $n_{\rm H}$, the recombination timescale of H\,II, which controls the availability of free electrons, is also shorter than the estimated duration of the SMBH activity. Thus, the steady-state assumption approximately holds.

\begin{figure*}
    \centering 
    \includegraphics[width=\textwidth]{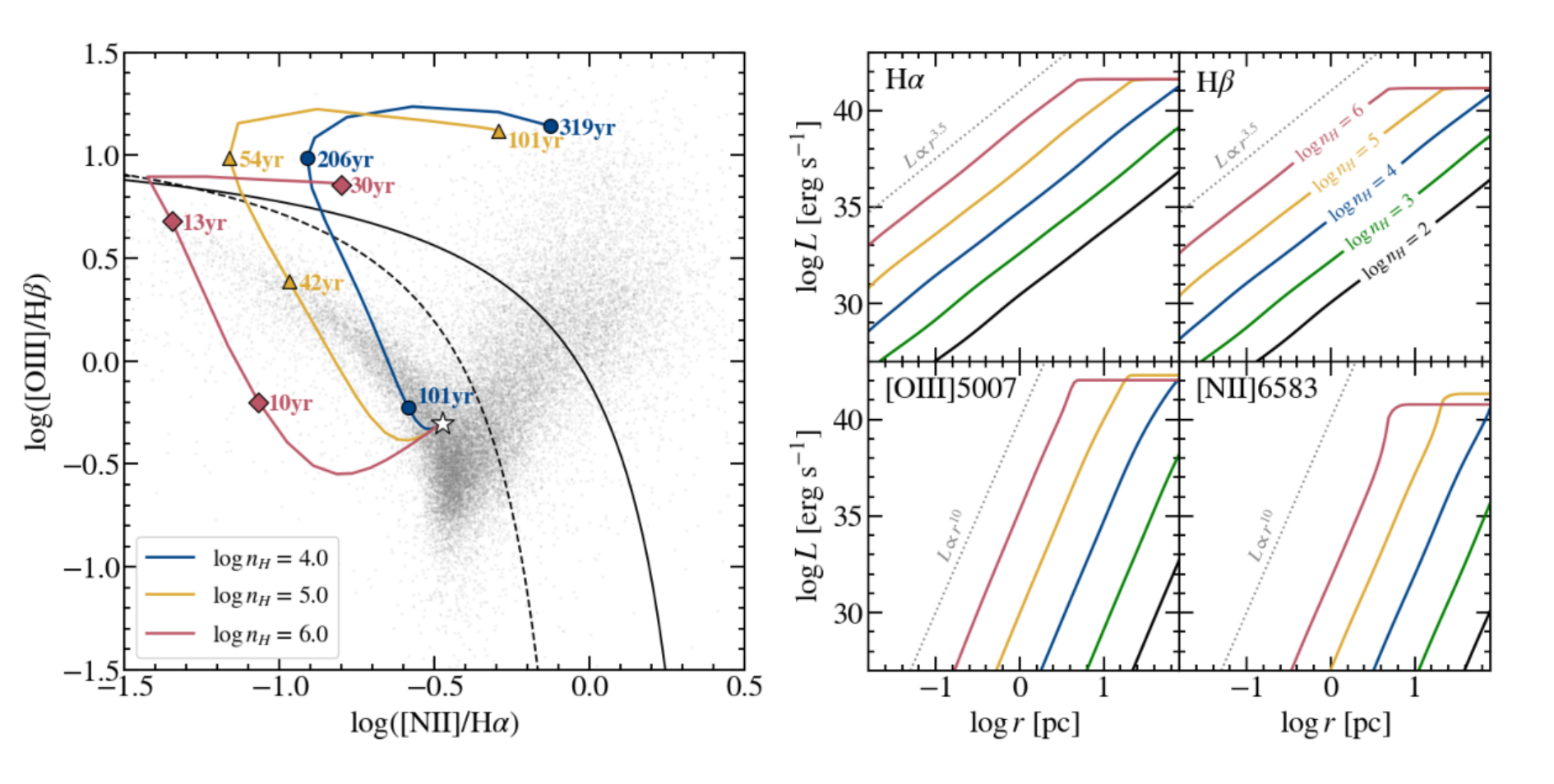}
    \caption[]{Examples of possible NLR evolution via \textsc{Cloudy} simulations at different gas densities. \textbf{Left:} The time-evolution of the source on a BPT diagram for $n_{\rm H}=10^4\,\mathrm{cm^{-3}}$, $n_{\rm H}=10^{5}\,\mathrm{cm^{-3}}$, $n_{\rm H}=10^{6}\,\mathrm{cm^{-3}}$. The white star marks the initial background line ratios from the assumed star-forming galaxy. We do not include the lower-density tracks because the total line luminosities are still dominated by the star-forming H\,II regions at $\sim100\,\mathrm{pc}$. The grey points are galaxies from the MPA-JHU DR7 catalogue\protect\footnotemark. \textbf{Right:} The line luminosity of selected optical lines as a function of the outer radius set in \textsc{Cloudy}. As discussed in the text, the radius is equivalent to time since the accretion system turned on ($t=r/c$). }
    \label{fig:time_evolution_bpt_line_luminosity}
\end{figure*}

In order to put another constraint on the line luminosities and to visualise the timescale on which the narrow-line region (NLR) emission in GSN\,069 evolves, we conducted another set of \textsc{Cloudy} simulations for $\log n_\mathrm{H}/\mathrm{cm^{-3}}=\{2,3,4,5,6\}$, assuming a fiducial volume filling factor of $f_\mathrm{V}=10^{-4}$. 
We set the inner radius of the gas to be $R_\mathrm{in}=0.010\,\mathrm{pc}$ and varied the outer radius between $R_\mathrm{out}=5\times10^{16}\,\mathrm{cm}\approx0.016\,\mathrm{pc}$ and $R_\mathrm{out}=3\times10^{20}\,\mathrm{cm}\approx100\,\mathrm{pc}$ since we are concerned about the integrated light from the entire NLR. An example of the \textsc{Cloudy} settings is provided in the Appendix. We present the results of these simulations in Figure \ref{fig:time_evolution_bpt_line_luminosity}. For the NLR evolution on the Baldwin-Phillips-Terlevich (BPT) diagnostic diagram \citep{BPT1981}, the starting point is a typical H\,II-dominated galaxy, which we assume has $L_\mathrm{H\alpha}=3\times10^{40}\,\mathrm{erg\,s^{-1}}$, $L_\mathrm{H\beta}= 10^{40}\,\mathrm{erg\,s^{-1}}$, $L_\mathrm{[O\,III]5007}=5\times10^{39}\,\mathrm{erg\,s^{-1}}$, and $L_\mathrm{[N\,II]6583}=10^{40}\,\mathrm{erg\,s^{-1}}$. All of the line ratios are computed by summing luminosities from both the assumed H\,II background and the simulated NLR. For the allowed gas densities and volume filling factor, the Str\"omgren sphere will expand approximately as $R = ct$ --- the standard isothermal Str\"omgren sphere expansion calculation would yield an unphysical superluminal expansion rate.
Therefore, different outer radii used in the simulation can be converted into time since the turn-on of the central ionising source. We see that initially the NLR is in the purely star-forming H\,II region in the BPT diagram, but as the Str\"omgren sphere expands over time, the NLR moves first to the upper-left corner and then transitions to moving rightward at nearly constant $\log{([\rm O\,III]/H_{\beta})}$. Eventually, the NLR settles in the AGN region of the BPT diagram after $\sim 30$--300\,yr depending on the gas density. The evolution is the fastest for higher $n_{\rm H}$. 

In our simulation, we found that the [O\,III] and [N\,II] line luminosities increase rapidly with the size of the Str\"omgren sphere, approximately scaling as $L \propto r^{10}$. Such rapid rise of [O\,III] luminosity can also be inferred from Figure \ref{fig:emissivity_Oiii}, since [O\,III] emissivity increases as $j_{\rm [O\,III]} \propto r^{7}$ for a constant $n_{\rm H}$ and it follows from Equation \ref{lum_emissivity} that $L_{\rm [O\,III]} \propto r^{10}$. On the other hand, the luminosities of the Balmer lines H$_{\alpha}$ and H$_{\beta}$ only scale as $\propto r^{3.5}$. The line luminosities remain relatively constant once the Str\"omgren sphere has expanded to a certain radius determined by the gas number density. This occurs because the emissivity of the lines diminishes significantly beyond this distance from the accretion system, making only a minor contribution to the overall integrated line luminosity. The point at which the line luminosities stabilise marks the transition of the NLR toward the AGN region in the BPT diagram. The implications are discussed in the following section.


\footnotetext{\hyperlink{https://wwwmpa.mpa-garching.mpg.de/SDSS/}{https://wwwmpa.mpa-garching.mpg.de/SDSS/}}


\section{Discussion and Prediction for Future Observations}
\label{sec:discussion}








Past X-ray observations show that the SMBH in GSN\,069 has been active for at least 13\,yr. An upper limit to the age of the accretion disk may be obtained from the following two arguments. First, the upper limit on the nuclear [O\,III] emission size is 35\,pc, which corresponds to a light-travel time of 115\,yr. A potential loophole in accepting this value as the upper limit to the age is that the size of the nuclear [O\,III] emitting region might simply be due to a high concentration of dense gas within 35\,pc. Second, in most QPE models (referred to in \S~I), the system is powered by a stellar-mass object orbiting an SMBH. Taking a radiative efficiency of $\eta\approx 0.1$, a mass budget of $M\approx 1\,M_\odot$, and an average bolometric luminosity of $L\approx 3\times10^{43}\rm\,erg\,s^{-1}$, we infer the lifetime of the system to be
\begin{equation}
    t_{\rm life} \approx \frac{\eta M c^2}{L}\approx 200\mathrm{\,yr} \frac{(\eta/0.1)\, (M/M_\odot)}{L/3\times10^{43}\rm\,erg\,s^{-1}}.
\end{equation}
From these two arguments, we infer that the current age of the system is likely less than $\mathrm{O}(10^2)\rm\, yr$. It is worth noting that {\it ROSAT} nondetection of bright X-ray emission in 1994 could mean that the SMBH is active for no more than $\sim 30$\,yr.

In Figure \ref{fig:contfree}, we see the presence of extended [O\,III] emission, albeit at a lower surface intensity than the nuclear [O\,III] region, stretching up to 2\,kpc away from the nucleus. This extended [O\,III] emission's origin could potentially be attributed to a previous phase of AGN activity, followed by the onset of bright X-ray emission sometime within the last 10--100\,yr. Indeed, \citet{Wevers_2022} have shown that, based on line fluxes from the entire galaxy, the ionising source in GSN\,069 is consistent with an AGN.


It is worth considering the possibility that a star-forming region within the nucleus could dominate as the source of the ionising radiation over the accretion disk. We can refute this possibility for the following reason. Let us consider an optimistic case of a young nuclear star cluster of mass $10^{6} \,M_{\odot}$ (a higher stellar mass will overproduce the far-UV fluxes in the nuclear region; see Table \ref{tab:photometry_all}). Employing the Salpeter initial mass function \citep{Salpeter_1955} results in an estimation of $\sim 10^{4}$ high-mass stars ($M \gtrsim 15\,M_{\odot}$). Each high-mass star emits $\mathcal{O}(10^{48})$ ionising photons ($h\nu>13.6$\,eV) per second \citep[see Table 15.1 of][]{Draine_2011_ISM_book}, resulting in emission of $\mathcal{O}(10^{52})$ ionising photons per second from the nuclear star cluster. In fact, the optimistic case of a young massive cluster we are considering here is too blue to be consistent with the observed far-UV to optical colour, which has a nearly flat SED with $\lambda F_\lambda \propto \lambda^{\approx 0}$ indicating an older stellar population with far fewer ionising photons. Comparatively, the accretion-disk SED model that is consistent with the X-ray observations \citep{Miniutti_2023A&A} produces roughly $\mathcal{O}(10^{53})$ ionising photons every second.
Thus, the budget of ionising photons is dominated by the accretion disk and not the nuclear stellar cluster. This rules out the scenario wherein the nuclear [O\,III] emission is powered by a nuclear star cluster.

Our model of an SMBH accretion disk that was recently activated by a sudden onset of gas supply in the past 10--100\,yr is consistent with all observations. To further test this picture, we consider the possible future evolution of the system. For simplicity, let us assume that the disk luminosity and SED stays more-or-less steady in the next few years. Based on the relatively modest H$\alpha$ luminosity from the compact nuclear region ($\ll 6\times10^{40}\rm\,erg\,s^{-1}$) and also the fact that soft X-rays are escaping from the system, we know that only a small fraction of the ionising, soft-X-ray photons from the accretion disk is used to power the compact nuclear [O\,III] region. Therefore, we expect that Str{\"o}mgren sphere surrounding the SMBH to expand at the speed of light, $\mathrm{d} r/\mathrm{d} t = c$. 
From Figure \ref{fig:time_evolution_bpt_line_luminosity}, it is clear that two distinct regimes exist for the evolution of [O\,III] luminosity. If the QPE is sufficiently young (i.e., before the luminosity flattens out), then we expect the [O\,III] luminosity to increase rapidly as $L \propto r^{10}$, with a fractional change given, to an order of magnitude, by 
\begin{equation}
    \frac{\diff L_{\rm [O\,III]}}{L_{\rm [O\,III]}} \approx \frac{10\diff r}{r} \approx \frac{10\diff t}{t}. 
\end{equation}
For a fixed time interval $\diff t$, we see that the [O\,III] luminosity will be less variable if the source is older. For instance, for $t\approx 30$\,yr and $\diff t=6$\,yr (for an observation taken in the near future), we expect a 200\% increase in the [O\,III] luminosity from the compact nuclear region. Such a large increase in the nuclear [O\,III] luminosity will be easily detectable by {\it HST}, and may also be observable with ground-based spectroscopic observations at low spatial resolutions, with the caveat that the measured line flux may be dominated by the extended [O\,III] region. Our power-law + point-source model shows that surface flux density declines with a rather shallow power law, $\mu (r) \propto r^{-1}$. It follows that the flux density enclosed within some radius $r$ scales as $F_{\lambda} (<r) \approx r^2\mu(r) \propto r$. This means that the unresolved [O\,III] region contributes only a small fraction of the total [O\,III] luminosity measured by ground-based observatories employing wider slits. For a $0.7''$-wide slit, the compact nuclear region only contributes $\sim 10\%$ of the [O\,III] luminosity. Thus, a 200\% increase in the [O\,III] luminosity from the NLR will translate to only a 25\% increase in total integrated flux, which may still be detectable by low-spatial-resolution spectroscopy.

The other regime is where the Str{\"o}mgren sphere has expanded beyond the point where the [O\,III] line luminosity levels off. In this situation, we will not observe any significant changes in the [O\,III] line's brightness. If this happens, we should also notice that the NLR has settled within the AGN region on the BPT diagram. In this case, the age of the QPE is degenerate with the gas number density ($n_{\rm H}$), but we can place even stronger constraints on both of these quantities. These arguments strongly support the idea of conducting similar {\it HST} observations, like the ones presented in this study, in the coming years to refine our understanding of the age of the SMBH accretion system in GSN\,069. Integral-field-unit spectra of GSN\,069 and its nucleus would be even more valuable in this context.

\section{Conclusion}
\label{sec:conclusion}
This paper presents an analysis of archival {\it HST} images, targeting the narrow-line [O\,III] emission within the QPE discovery source, GSN\,069. With aperture photometry on the continuum-subtracted [O\,III] image, we find strong evidence for the presence of a compact, unresolved nuclear [O\,III] emission region, which is overlaid on top of a spatially extended [O\,III] emission region reaching as far as 2\,kpc away from the nucleus. The extended [O\,III] emission does not correspond significantly with star-forming H\,II regions, making it likely that the extended emission is due to past AGN activity instead of ionisation by hot stars.

We measured the [O\,III] luminosity of the compact region, yielding a value of $(2.1 \pm 0.3) \times 10^{40}\,\rm erg\,s^{-1}$. Using \textsc{Cloudy} simulations, we constrain the physical properties of the [O\,III]-emitting gas in the compact region, and our findings are summarised as follows:
\begin{enumerate}
    \item size of emitting region $4\,\text{pc} < r < 35\,\text{pc}$,
    \item age of the SMBH accretion system $10\lesssim t \lesssim 100$\,yr,
    \item gas number density $5 \times 10^{3} < n_{\rm H} \lesssim 10^{8}\, \rm cm^{-3}$,
    \item volume filling factor $f_{\rm V} < 2 \times 10^{-3}$, and
    \item mass of the emitting gas $M_{\text{gas}} > 3 \times 10^{3}\,M_{\odot}$.
\end{enumerate}
These constraints lead us to conclude that the [O\,III] emission in the compact region comes from dense gas that likely originates from dense molecular clouds within the nuclear region.  Such molecular clouds also exist at the Galactic Center of our Milky Way Galaxy. Compared to other known AGNs whose long-lasting radiation pressure has largely modified its gaseous environment, the nucleus of GSN\,069 is a unique, young (most likely 10 to 100\,yr-old) ionising source that allows us to probe the pristine gas environment near the otherwise inactive SMBH. This likely applies to other QPE sources as well. Finally, based on our current understanding, we predict that the [O\,III] luminosity from the compact region will likely undergo a $\sim 200\%$ change in a time interval of a few years if the QPE is young. If the line luminosity increases, we should also see that the NLR lies in the H\,II region of the BPT diagram. On the other hand, if the QPE is older (or the gas in the NLR is denser), we will not see any change in [O\,III] line luminosity. In this case, the NLR will be found in the AGN region of the BPT diagram.
This strongly motivates another epoch of {\it HST} observations similar to the ones presented here in the near future, which will further pin down the age of the SMBH accreting system.

\section*{Acknowledgements}
Research by K.C.P. is funded in part by generous support from Sunil Nagaraj, Landon Noll, and Sandy Otellini. We acknowledge the support from the Rose Hills Innovator Program. G.M. was supported by grant PID2020-115325GB-C31 funded by MICIN/AEI/10.13039/501100011033. A.V.F. is grateful for support from the Christopher R. Redlich Fund and numerous other donors.  This research benefited from interactions at workshops funded by the Gordon and Betty Moore Foundation through frant GBMF5076.

\section*{Data Availability}
The {\it HST} images used in this work are publicly available on STScI's MAST.

\section*{Appendix}\label{sec:appendix}

An example of the \textsc{Cloudy} input settings used to create Figure \ref{fig:emissivity_Oiii} is shown below for reference. 

\begin{lstlisting}
luminosity total 43.48
table SED "disk.sed" 
radius 18.70
hden 2.48
abundances "default.abn" no grains
iterate to converge
print last iteration
stop thickness 11.00
save lines, emissivity, "R18.7nH2.48.lem" last
blnd 5007.00A 
h 1 6562.80A 
h 1 4861.33A 
end of lines 
\end{lstlisting}

Below we provide an example of the \textsc{Cloudy} settings used to create Figure \ref{fig:time_evolution_bpt_line_luminosity}:
\begin{lstlisting}
luminosity total 43.48
table SED "shortdisksed.sed" 
filling factor 0.0001 
radius 16.50 19.75
hden 3.00
abundances "default.abn" no grains
iterate to converge
print last iteration
save lines, list"R19.75nH3.0fV1e-4.lines" 
    "LineList_NLR.dat" 
    units Angstrom 
    absolute last
save lines, emissivity"R19.75nH3.0fV1e-4.lem"
    units Angstrom last
o 3 5006.84A 
blnd 5007.00A 
h 1 6562.80A 
h 1 4861.33A 
n 2 6583.45A 
s 2 6716.44A 
s 2 6730.82A 
o 1 6300.30A 
end of lines 
save overview "GSN069_R16.7nH3.0fV1e-4.ovr" last
save continuum "GSN069_R16.7nH3.0fV1e-4.con" 
    units Angstrom last
\end{lstlisting}



\bibliographystyle{mnras}
\bibliography{references.bib} 







\bsp	
\label{lastpage}
\end{document}